\documentstyle[12pt]{oshiro}

\begin{document}



\begin{flushright}
  DPNU-93-31 \\
  August 1993
\end{flushright}
 
    \baselineskip 2pc

\vspace*{1.5cm}
    {\large
      \begin{center}
        {\bf Quantum Fluctuations of Black Hole Geometry}
      \end{center}
  \vspace{0.5cm}
  
      \begin{center}
      {\bf Kouji NAKAMURA\footnote{e-mail:kouchan@allegro.phys.nagoya-u.ac.jp},
        Shigelu KONNO, Yoshimi OSHIRO } \\
        and {\bf Akira TOMIMATSU}\\
      \end{center}
    }
  \vspace{0.5cm}
  
      \begin{center}
        {\it Department of Physics, Nagoya University\\
          Chikusa-ku, Nagoya 464-01,Japan}\\
      \end{center}
  \vspace{0.5cm}

        \hspace*{\parindent}
        By using the minisuperspace model for the interior metric of static 
        black holes, we solve the Wheeler-DeWitt equation to study quantum 
        mechanics of the horizon geometry.
        Our basic idea is to introduce the gravitational mass and the 
        expansions of null rays as quantum operators.
        Then, the exact wave function is found as a mass eigenstate, 
        and the radius of the apparent horizon is quantum-mechanically 
        defined.
        In the evolution of the metric variables, the wave function changes 
        from a WKB solution giving the classical trajectories to a tunneling 
        solution.
        By virtue of the quantum fluctuations of the metric evolution beyond 
        the WKB approximation, we can observe a static black hole state with 
        the apparent horizon separating from the event horizon.

\newpage
\section{Introduction} 
    \label{sec:int}

        \hspace*{\parindent}
        Since the discovery of the Hawking radiation, much work has been 
        devoted to the analysis of the quantum evaporation of black holes. 
        In particular, the problem of final fates of evaporating black holes 
        has been much debated. 
        In Hawking's semiclassical calculation, the emitted radiation was 
        found to be exactly thermal. 
        Then, if a black hole evaporates completely, an initially pure quantum 
        state must evolve to a mixed state. 
        This is known as the information loss paradox. 
        As was emphasized by Preskill\cite{Preskill}, it is very difficult to 
        resolve the serious puzzle in quantum mechanics and general relativity.
        Before reaching the final resolution, we must develop quantum theories 
        of the black hole geometry.
        
        A possible way of treating the horizon as a quantum system is to apply 
        the Wheeler-DeWitt equation to spherically symmetric spacetimes. 
        In the superspace canonical formulation, the Hamiltonian and momentum 
        constraints work as quantum equations for the physical state $\Psi$ 
        which is a functional of the metric variables. 
        To make the calculation tractable, Rodrigues et al\cite{Rodrigues} 
        proposed a black hole minisuperspace model and derived the simplified 
        Hamiltonian constraint. 
        Unfortunately their model was found to be incompatible with the 
        momentum constraint\cite{Lousto}. 
        The compatibility between the two constraints can be recovered if we 
        consider a local minisuperspace model valid near the apparent 
        horizon\cite{ATomimatsu}. 
        From the  wave function dependent on the local metric near the 
        apparent horizon, we can derive the mass-loss rate due to the 
        back-reaction of Hawking radiation and show the breakdown of the 
        semi-classical result at the final stage of complete evaporation. 
        The Wheeler-DeWitt approach will be viable as a quantum theory of the 
        horizon. 
        To advance this prospect, in this paper, we want to clarify another 
        quantum feature of the horizon from the Wheeler-DeWitt equation. 
        
        We will consider static states of a spherically symmetric black hole 
        instead of its evolutionary states. 
        In classical relativity the apparent horizon is always located just on 
        the event horizon. 
        The degeneracy of the two horizons can be removed owing to quantum 
        fluctuations of the metric. 
        This was first pointed out by York\cite{York} under the semiclassical 
        approximation (which treats the time-averaged vacuum Eisntein 
        equations), and the notion of ``quantum ergosphere" was introduced 
        to explain the origin of black hole radiance. 
        The purpose of our work is to give a Wheeler-DeWitt description for 
        the interesting quantum phenomenon. 
        
        For any static, spherically symmetric spacetimes we can choose the 
        metric dependent on a radial coordinate $r$ only. 
        Note that if we are concerned with the interior geometry of black 
        holes, the coordinate $r$ plays the role of a time coordinate, and we 
        have a time slicing on the homogeneous spatial hypersurface $r=$const. 
        Because of this homogeneity the momentum constraint becomes trivial, 
        and the Hamiltonian constraint reduces to a partial differential 
        equation. 
        
        In \ref{sec:Basic} we construct the canonical variables from the 
        metric for the interior geometry and write down the Hamiltonian. 
        The spherically symmetric metric permits us to introduce the 
        locally-defined gravitational mass\cite{FMP} and the expansion of null 
        geodesics, which are represented by the canonical variables. 
        The mass is required to be conserved for the spherically symmetric 
        Ricci flat system, and the expansion of null geodesics is used to 
        determine the radius of the apparent horizon. 
        Our key idea is to treat these geometrical quantities as 
        quantum-mechanical operators. 
        Nambu and Sasaki\cite{NambuSasaki} discussed the WKB solutions of the 
        Wheeler-DeWitt equation by using classical trajectories in the 
        minisuperspace. 
        In \ref{sec:quantum} we specify the exact solution representing 
        the interior of the static hole as an eigenstate of the mass operator. 
        (The Appendix contains some other examples of the mass eigenstate.) 

        From the wave function, in \ref{sec:fluct}, we discuss the quantum 
        separation of the apparent and event horizons. 
        In the minisuperspace of the metric variables we find the classically 
        forbidden region between the two horizons where the classical action 
        becomes imaginary.
        It is shown that the Wheeler-DeWitt equation can give a plausible wave 
        function to model quantum fluctuations of the horizon geometry. 
        
        In this paper, we denote the Planck constant and the Newton's 
        gravitational constant by $\hbar$ and $G$, respectively, and  we use 
        units such that the light velocity $c=1$.

\section{Quantum Operators}
    \label{sec:Basic}

        \hspace*{\parindent}
        We follow the standard canonical formulation of general relativity, by 
        choosing a time slicing. 
        The Einstein Lagrangian density is decomposed into the extrinsic and 
        intrinsic curvatures of a spatial hypersurface. 
        To construct the Hamiltonian and momentum constraints, we give the 
        canonical momentum conjugate to the 3-metric on the hypersurface. 
        
        Let us consider the extended Schwarzschild spacetime with the metric 
        written in the form 
    \begin{equation}
          \label{eqn:metric}
         ds^{2} = - \frac{\alpha^{2}}{u} dT^{2} + u dX^{2} 
                  + v (d\theta^{2} + \sin^{2}\theta d\phi^{2}),
    \end{equation}
        where $\alpha$ is the lapse function.
        A Cauchy surface of the whole spacetime may be chosen as a 3-surface 
        $T=$const., which is the solid line drawn in Fig.1.
        On the straight line segment (A) of the hypersurface, $T$ should be 
        regarded as the Kruskal time-coordinate, and the metric coefficients 
        $u$ and $v$ will depend on both $T$ and $X$.
        However, on the curved line segment (B), $T$ can be identical with the 
        Schwarzschild spherical coordinate $r$ which works as a 
        time-coordinate inside the black hole, and under the gauge choice 
        $\alpha=1$ the metric given by 
    \begin{equation}
          \label{eqn:Schwarzschildmetric}
        u = \frac{2Gm}{T} - 1, \; \; \; \; v=T^{2}
    \end{equation}
        becomes independent of the coordinate $X$.
        In this paper we will treat the metric fluctuations on the homogeneous 
        part of the Cauchy surface (the curved line segment (B)), by assuming 
        the Kantowski-Sachs minisurperspace model such that 
        $\alpha=\alpha(T), u=u(T), v=v(T)$.
        
        The Einstein Lagrangian for this model inside the black hole reduces 
        to 
    \begin{equation}
          \label{eqn:lagrangian}
       L  =  \frac{V}{4 G} ( - \frac{\dot{v} \dot{u}}{\alpha}
        - \frac{u \dot{v}^{2}}{2 \alpha v} + 2 \alpha ),
    \end{equation}
        where the dots denote derivatives with respect to the coordinate $T$, 
        and the length $V = \int_{(B)}dX$ of the line segment (B) is assumed 
        to be a finite constant.
        From the Lagrangian (\ref{eqn:lagrangian}) we obtain the canonical 
        momenta conjugate to the variables $u$ and $v$
    \begin{equation}
         \label{eqn:momenta}
      \Pi_{u} = - \frac{V}{4 G} \frac{\dot{v}}{\alpha} , \mbox{\ \ \ \ } 
      \Pi_{v} = - \frac{V}{4 G} (\frac{\dot{u}}{\alpha} + 
      \frac{u \dot{v}}{\alpha v} ), 
    \end{equation}
        and the Hamiltonian written in the form
    \begin{eqnarray}
        H & = & \Pi_{u} \dot{u} + \Pi_{v} \dot{v} - L \nonumber \\
          \label{eqn:hamiltonian}
          & = & \frac{4 G \alpha}{V} \left[ - \Pi_{u} \Pi_{v} + 
                \frac{u \Pi_{u}^{2}}{2 v} - \frac{V^{2}}{8 G^{2}} \right], 
     \end{eqnarray}
        which gives the dynamical constraint $H = 0$. 
        
        To quantize the system, we make the usual substitutions for the 
        momenta 
    \begin{equation}
          \Pi_{u}\rightarrow \frac{\hbar}{i}\frac{\partial}{\partial u}, 
          \; \; \; \;
          \Pi_{v}\rightarrow \frac{\hbar}{i}\frac{\partial}{\partial v}. 
          \label{eqn:momentamsubstitutions}
    \end{equation}          
        Then the quantum state of the black hole interior is represented by 
        the wave function $\Psi(u,v)$ on the minisuperspace, which satisfies 
        the Wheeler-DeWitt equation
    \begin{equation}
         H \Psi = 0.
      \label{eqn:Wheeler-DeWitt}
    \end{equation}
        
        To discuss quantum mechanics of the horizon geometry from the wave 
        function, we introduce some geometrical quantum operators.
        For the spherically symmetric metric satisfying the vacuum Einstein 
        equations, we have a locally defined gravitational mass $M$ as a 
        dynamical constant.
        If we use the canonical momenta (\ref{eqn:momenta}), $M$ has the form
    \begin{equation}
         \label{eqn:massfunction}
      M = \frac{2 G u \Pi_{u}^{2}}{V^{2} v^{1/2}}  + \frac{v^{1/2}}{2 G}.
    \end{equation}
        which is weakly commutable with the Hamiltonian as follows,
    \begin{equation}
         \label{eqn:hamiltonianandmasscommutation}
      [ H , M ] \Psi = - \frac{2 i G \hbar \Pi_{u}}{V^{2} v^{1/2}} H \Psi = 0.
    \end{equation}
        This commutation relation means that the physical state $\Psi$ can be 
        a mass eigenstate.
        
        Inside the black hole there exists a trapped region bounded by the 
        apparent horizon, which is defined in terms of the expansion of null 
        geodesics.
        Following Carter\cite{Carter}, we consider a null-vector decomposition 
        of the metric (\ref{eqn:metric}),
    \begin{equation}
         \label{eqn:metricdecomposition}
           g_{ab} = - \beta_{a} l_{b} - \beta_{b} l_{a} + \gamma_{ab},
    \end{equation}
        with
    \begin{eqnarray}
          l^{a}l_{a} = \beta^{a} \beta_{a} = 0, \mbox{\ \ \ \ }
          l^{a} \beta_{a} = -1, \nonumber \\
             \label{eqn:nullcongruencesconditions}
          l^{a} \gamma_{ab} = \beta^{a} \gamma_{ab} = 0, 
    \end{eqnarray}
        where $l^{a}$ and $\beta^{a}$ are the vector fields tangent to 
        outgoing and incoming null rays, respectively.
        In any spherical symmetric system, these two null-vectors are 
        necessarily tanget to null geodesics.
        Then, from the expansions for outgoing and ingoing null rays given by 
    \begin{equation}
        \label{eqn:nullexpansion}
       \theta_{+} = l^{a}_{\; \; ;a} + \beta^{a} l^{b} l_{a;b}, \mbox{\ \ \ \ }
       \theta_{-} = \beta^{a}_{\; \; ;a} + \l^{a} \beta^{b} \beta_{a;b}, 
    \end{equation}
        we obtain the useful relation
    \begin{equation}
        \label{eqn:expansion}
       \theta_{-} \theta_{+} = \frac{8 G^{2} u \Pi_{u}^{2}}{ V^{2} v^{2}}.
    \end{equation}
        The substitution of Eq.(\ref{eqn:massfunction}) leads to 
    \begin{equation}
         \label{eqn:massexpansionrelation}
       \theta_{-}\theta_{+} = - \frac{2}{v}\left(1-\frac{2GM}{v^{1/2}}\right)
    \end{equation}
        In classical geometry, both $\theta_{+}$ and $\theta_{-}$ become 
        negative in the trapped region, and $\theta_{+}=0$ on the apparent 
        horizon.
        The event horizon will be the null surface $u=0$, where in the 
        Schwarzschild spacetime the expansion $\theta_{+}$ also vanishes.
        In the quantum version we must treat $\theta_{-}\theta_{+}$ as an 
        operator.
        However, the canonical quantum theory of gravity gives no general 
        procedure for extracting geometrical informations of the spacetime 
        from the wave function.
        Fortunately, this obstacle can be partly overcome if we consider a 
        static black hole state corresponding to an eigenstate of the mass 
        opeator $M$;
    \begin{equation}
        \label{eqn:massegenfunction}
           M \Psi = m \Psi,
    \end{equation}
        where $m$ is the mass eigenvalue and remains constant in the whole 
        spacetime.
        In our minisuperspace model, the metric variables $u$ and $v$ will be 
        observables on each homogeneous spatial hypersurface $T=$const.
        (Here we assume that dynamical variables which are invariant under 
        spatial coordinate transformations are true observalbes, even if they 
        do not commute with the Hamiltonian constraint\cite{Kuchar}.
        Such dynamical variables can be the 3-Ricci scalar ${}^{(3)}R=2/v$ 
        and 3-volume $4\pi V \sqrt{u} v$ of a hypersurface $T=$const.)
        Then, for the mass eigenstate $\Psi$, 
        Eq.(\ref{eqn:massexpansionrelation}) assures that 
        $\theta_{-}\theta_{+}$ also is a observable on each hypersurface. 
        If the condition $\Psi^{*}\theta_{-}\theta_{+}\Psi>0$ 
        (i.e., $v^{1/2}<2Gm$) is satisfied on a hypersurface, it will mean a 
        trapped surface. 
        The apparent horizon will be just the hypersurface where 
        $\Psi^{*}\theta_{-}\theta_{+}\Psi=0$. 
        ( Note that each hypersurface is topologically $R\times S^{2}$. 
        The variable $v^{1/2}$ denotes a proper radius of $S^{2}$ and 
        characterizes the position of the hypersurface in the spacetime.) 
        In general, the event horizon can be defined as a null surface which 
        has a finite proper radius of $S^{2}$.
        This definition will be valid in quantum geometry, in which $u$ and 
        $v$ become observables.
        We can observe the null surface $u=0$ among various $T=$const. 
        hypersurfaces with finite $v$ as the event horizon.
        Thus, in the minisuperspace with the coordinates $u$ and $v$, we 
        require the positions of the apparent and event horizons to be 
        $v^{1/2}=2Gm$ and $u=0$ respectively. 
        The important point is that the wave function allows us to observe 
        a nonvanishing value of the variable $u$ on the hypersurface 
        $v^{1/2}=2Gm$ as a result of quantum separation of the two horizons. 
        In fact, Eq.(\ref{eqn:massfunction}) means that $u$ may not vanish on 
        the hypersurface $v^{1/2}=2Gm$ by the virtue of the uncertainty 
        principle $\Delta u \cdot \Delta \Pi_{u} \sim \hbar$.

\section{Wave Function}
    \label{sec:quantum}

        \hspace*{\parindent}
        For the canonical quantization of the theory, we meet with the problem 
        of operator ordering. 
        Taking account of this ambiguity, we write the Hamiltonian operator in 
        the form
    \begin{equation}
       \label{eqn:Hamiltonian operator}
        H = \frac{4 G \alpha}{V} \left( - \Pi_{u} \Pi_{v} + 
                  \frac{u^{p} \Pi_{u} u^{1-p} \Pi_{u}}{2 v} - 
                  \frac{V^{2}}{8 G^{2}} \right).
    \end{equation}
        Then, to keep the commutation relation 
        (\ref{eqn:hamiltonianandmasscommutation}), the mass operator must be 
    \begin{equation}
         \label{eqn:massoperater}
      M \Psi = \left(\frac{2 G u^{p} \Pi_{u} u^{1-p} \Pi_{u}}{V^{2} v^{1/2}} 
                                            + \frac{v^{1/2}}{2 G} \right) \Psi.
    \end{equation}
        The Wheeler-DeWitt equation $H\Psi=0$ supplemented by the condition 
        (\ref{eqn:massegenfunction}) of the mass eigenstate gives the unique 
        exact solution
    \begin{equation}
          \label{eqn:schwarzshildwavefunction}
       \Psi = N \frac{z^{p}}{(v - 2 G m v^{1/2})^{p}} H_{p}^{(1)}(z),
    \end{equation}
        where $N$ is an arbitrary constant, and
    \begin{equation}
          \label{eqn:argument}
        z = \frac{V}{G \hbar} \sqrt{- u (v - 2 G m v^{1/2})}. 
    \end{equation}
        We can obtain similar solutions for quantum extensions of the 
        Reissner-Nordstr\"om and Schwarzschild-de Sitter metrics 
        (see Appendix).
        The choice of $H^{(1)}_{p}(z)$ instead of $H^{(2)}_{p}(z)$ is 
        verified because for the WKB approximation valid in the region $z>1$ 
        the wave function behaves as 
    \begin{equation}
         \label{eqn:wkbapproximationofexactsolution}
       \Psi \sim e^{i z}, 
    \end{equation}
        where $\hbar z$ corresponds to the classical action $S$ of the form
    \begin{equation}
         \label{eqn:z=classicalaction}
       S = \frac{V}{G}\sqrt{ - u (v - 2 G m v^{1/2}) }. 
    \end{equation}
        This classical action corresponds to one obtained by Nambu and Sasaki
        \cite{NambuSasaki}, and by Fischler et al\cite{FMP}. 
        
        Because the Hamilton-Jacobi theory requires
    \begin{equation}
             \label{eqn:hamiltonjacobirelation}
      \Pi_{u} = \frac{\partial S}{\partial u}, \; \; \; \; 
      \Pi_{v} = \frac{\partial S}{\partial v},
    \end{equation}
        from Eqs.(\ref{eqn:momenta}) and (\ref{eqn:hamiltonjacobirelation}) 
        the WKB wave function gives the trajectories of classical solutions 
        on the ($u,v^{1/2}$)-plane as follows,
    \begin{equation}
            \label{eqn:classicaltrajectory}
        u = - c \left( 1 - \frac{r_{g}}{v^{1/2}} \right),
    \end{equation}
        where $c$ is the positive constant of integration, and $r_{g}=2Gm$.
        The classical trajectories with different $c$ represent only one 
        physical trajectory, since the constant $c$ appears owing to the 
        degree of freedom of the scale transformation of the coordinate $X$. 
        The action $S$ vanishes at the classical singlarity $v=0$, where the 
        WKB approximation should break down.
        This occurs also on the apparent horizon $r=r_{g}$ and the event 
        horizon $u=0$.
        Thus quantum evolutionary paths of the metric variables near the 
        horizons can deviate from Eq.(\ref{eqn:classicaltrajectory}) which is 
        derived from the semiclassical wave function 
        (\ref{eqn:schwarzshildwavefunction}).

\section{Separation of Two Horizons}
    \label{sec:fluct}

        \hspace*{\parindent}
        On the classical trajectories given by 
        Eq.(\ref{eqn:classicaltrajectory}) the action 
        (\ref{eqn:z=classicalaction}) remains real.
        However, the exact solution (\ref{eqn:schwarzshildwavefunction}) 
        permits us to consider the region where $\arg z$ is chosen to be 
        $\pi/2$.
        (We require that in the region where $z$ becomes a pure imaginary 
        $\Psi$ must exponentially decrease as $|z|$ increases.
        In the following discussions we need not specify the values of the 
        non-WKB terms, such as $z^{p}$, contained in $\Psi$.)
        Let us divide the ($u,v^{1/2}$)-plane into the following four regions 
        shown in Fig.2a: The region A and B where $u>0$ represent the metric 
        of the black hole interior, while the metric outside the event horizon 
        occupies the regions C and D where $u<0$. The metric on the trapped 
        surface where $\Psi^{*}\theta_{-}\theta_{+}\Psi>0$ must be a point in 
        the region A and C where $v^{1/2}<r_{g}$.
        
        Note that the boundary $v^{1/2}=r_{g}$ between A and B (or between C 
        and D) represents the metric on the apparent horizon where 
        $\Psi^{*}\theta_{-}\theta_{+}\Psi=0$. 
        In other words, $T=const.$ surface represented by $v^{1/2} = r_{g}$ 
        is the apparent horizon.
        Unless $u=0$ on the boundary, we have a geometrical structure of the 
        apparent horizon which is separated from the event horizon.
        Of course, the classical trajectories (\ref{eqn:classicaltrajectory}) 
        which cover the regions A and D always pass through the point given by 
        $u=0$ and $v^{1/2}=r_{g}$, where the two horizons are degenerate.
        In fact, for the mass eigenstate $\Psi$ we have the equation
    \begin{equation}
         u^{p}\Pi_{u}u^{1-p}\Pi_{u}\Psi=
             \frac{V^{2}v^{1/2}}{2G^{2}}(r_{g}-v^{1/2})\Psi,
          \label{eqn:upi2psi}
    \end{equation}
        (see Eq.(\ref{eqn:expansion})) which means the existence of a 
        potential barrier to make $\Pi_{u} \cong \dot{v}$ vanish on the 
        boundary $v^{1/2}=r_{g}$.
        Any wave packet with the oscillatory form 
        (\ref{eqn:wkbapproximationofexactsolution}) in the classically 
        allowed region A may not be able to propagate across the boundary, 
        except the gate opened at $u=0$. 
        However, when $z\leq1$ in the region A (see Fig.2b), the WKB 
        approximation breaks down, and quantum evolutionary paths of the 
        metric variables $u$ and $v$ will considerably fluctuate from the 
        classical ones. 
        This can be also seen from the fact that the wave function 
        (\ref{eqn:schwarzshildwavefunction}) becomes almost independent of 
        the variables $u$ in the limit $v^{1/2}\rightarrow r_{g}$. 
        Though the notion of probability is quite obscure, the 
        usual quantum-mechanical interpretation will reject the dominant 
        contribution of the black hole state with completely degenerate 
        horizons. 
        The equation (\ref{eqn:schwarzshildwavefunction}) rather describes the 
        quantum penetration of the wave function into the classical forbidden 
        region B. 
        Because $\Pi_{u}^{2}<0$ in the region B, the metric variable $v$ 
        evolves with the imaginary time. 
        If $|z|>1$, then wave function exponentially decreases.
        Therefore, to avoid the high potential barrier, this tunneling process 
        will occur through the limited region $|z|\leq 1$ and induce the 
        evolution of $u$ toward the boundary $u=0$ between the regions B and D.
        
        Recall that our canonical analysis is not applicable to the
        region C and D ($u<0$), because the $T$=const. hypersurface becomes
        timelike outside the event horizon. 
        Nevertheless, the analytical continuation of 
        Eq.(\ref{eqn:schwarzshildwavefunction}) to $u<0$ shows plausible 
        behaviors of the wave function: There exist the classical trajectories 
        (\ref{eqn:classicaltrajectory}) in the region D, and the WKB 
        approximation breaks down near the boundary $u=0$. 
        Though in this paper we have no precise analysis of the wave function 
        for the metric in the region C and D, we can claim that the wave 
        function (\ref{eqn:schwarzshildwavefunction}) gives a tunneling 
        solution (in the region B) connecting two oscillatory solutions 
        inside the apparent horizon and outside the event horizon. 
        This connection should occur through the restricted region $|z|\leq 1$ 
        near the two horizons.
        
        If one consider a tunneling process passing through the region
        C, the apparent horizon must be present outside the event horizon.
        This case will be more important in relation to the black hole
        radiance. 
        In York's semiclassical argument, the separation of the two horizons 
        is due to some mass fluctuations. To contain this effect, the wave 
        function must be constructed by a superposition of 
        Eq.(\ref{eqn:schwarzshildwavefunction}) with different masses. 
        Such a superposition will induce a spreading of the width of allowed
        classical trajectories, which need not pass through the point given by
        $u=0$ and $v^{1/2}=2G\bar{m}$ ($\bar{m}$ is the averaged mass). 
        Then, we can see the separation of the two horizons in the 
        semiclassical level.
        However, this semiclassical mechanism will be a result (a
        backreaction) of the Hawking radiation rather than its origin. 
        If the separation of the two horizons has some relevance to the black 
        hole radiance, we must explain the origin in terms of the quantum 
        tunneling process discussed here. 
        This point should be investigated in future works, which include a 
        more precise derivation of the wave function outside the event horizon.

        In summary, under the minisuperspace model with metric
        variables $u$ and $v$ of the black hole interior, we have found the
        exact wave function as a mass eigenstate, by introducing the quantum
        operators corresponding to the gravitational mass and the expansions
        of null rays. 
        The wave function describes a change from the WKB solution giving 
        classical trajectories to the tunneling solution representing the 
        separation of the apparent and event horizons. 
        The quantum fluctuations beyond the WKB approximation must become 
        important in the restricted region $|z|\leq 1$.
        For example, the classical trajectory given by $u=-(1-r_{g}/v^{1/2})$
        enters into the tunneling region, when the variable $v$ evolves into
        the range $|(v^{1/2}/r_{g})-1| \leq l_{p}^{2}/Vr_{g}$. 
        If the gravitational radius $r_{g}$ is much larger than the Planck 
        length $l_{p}$, the tunneling region remains very narrow, and the 
        separation of the two horizons becomes negligibly small 
        ($V \geq l_{p}$). 
        This fits well the standard point of view that any quantum effects 
        must be suppressed for large black holes. Thus we conclude that the
        Wheeler-DeWitt approach based on the canonical quantization of
        geometrical quantities is useful for quantum mechanics of the horizon.

\section*{Acknowledgements}

        \hspace*{\parindent}

        We thank to Y. Nambu for useful discussions. 
	This work is partially supported by the Grant-in-Aid for Scientific 
	Research of Ministry of Education, Science and Culture of Japan 
	(No. 04640268).

\appendix

\section*{Wave Functions of Mass Eigenstates}
    \label{sec:RissneranddeSitter}
       \setcounter{section}{1}
       \setcounter{equation}{0}

        \hspace*{\parindent}
        In the text we have considered the Wheeler-DeWitt equation constructed 
        form the Hamiltonian constraint for the vacuum Einstein equation, and 
        the wave function has been found as an eigenstate of the quantum mass 
        operator. 
        This idea can be applied to some other static systems of the 
        spherically symmetric metric. Here we give the wave functions 
        analogous to Eq.(\ref{eqn:schwarzshildwavefunction}) as quantum 
        extensions of the Reissner-Nordstr\"om and Schwarzschild-de Sitter 
        metrics.
        
        First, let us add the term 
    \begin{equation}
          \label{eqn:emterm}
       L_{A} = \frac{V v \dot{A}^{2}}{2 \alpha}
    \end{equation}
        to the Lagrangian (\ref{eqn:lagrangian}), where $A$ is the electric 
        potential dependent only on $T$ inside the black hole. Then we can 
        obtain the total Hamiltonian
    \begin{equation}
          \label{eqn:totalhamiltonian}
       H_{T} = H + \frac{\alpha \Pi_{A}^{2}}{2 V v},
    \end{equation}
        where $\Pi_{A}$ is the canonical momentum conjugate to $A$.
        Under the gauge choice $\alpha = 1$, the Reissner-Nordstr\"om metric 
        is given by 
    \begin{equation}
          \label{eqn:reisolution}
       u = - ( 1 - \frac{2 G m}{T} + \frac{Q^{2}}{T^{2}} ), 
                                     \mbox{\ \ \ \ } v = T^{2}.
    \end{equation}
        In the classical level, Hamiltonian constraint requires
    \begin{equation}
          \label{eqn:piaeigenvalue}
      \Pi_{A}^{2} = \frac{V^{2}}{G} Q^{2}.
    \end{equation}
        This can remains valid in the quantum level, because we have the 
        commutation relation 
    \begin{equation}
         \label{eqn:pia-hcommutationrelation}
       [ \Pi_{A}, H ] = 0.
    \end{equation}
        Further we can find the total mass operator $M_{T}$ defined by 
    \begin{equation}
          \label{eqn:rissnerbearmass}
       M_{T} = M + \frac{\Pi_{A}^{2}}{2 V^{2} v^{1/2}},
    \end{equation}
        which is weakly commutable with $H_{T}$ as follows, 
    \begin{equation}
          \label{eqn:bearmasshamiltoniancommutationretationofreissner}
       [ H_{T} , M_{T} ]
            = \frac{2 i G \hbar \Pi_{u}}{V^{2} v^{1/2}} H_{T}.
    \end{equation}
        These commutation relations assure the existence of the exact 
        solution with the eigenvalues of total mass $m$ and charge 
        $Q/\sqrt{G}$ for the Wheeler-DeWitt equation, and we obtain the wave 
        function 
    \begin{equation}
          \label{eqn:Reissner-Nordstromwavefunction}
      \Psi = N \frac{z^{p}}{(v - 2 G m v^{1/2} + Q^{2})^{p}} H_{p}^{(1)}(z) 
                e^{\pm i Q/\sqrt{G} \hbar},
    \end{equation}
        where 
    \begin{equation}
          \label{eqn:Reissner-Nordstromargument}
      z^{2} = - \frac{V^{2} u (v - 2 G m v^{1/2} + Q^{2})}{ G^{2} \hbar^{2}}.
    \end{equation}
        In the ($u,v^{1/2}$)-plane the separation of the apparent horizon 
        ($v-2Gmv^{1/2}+Q^{2}=0$) and the event horizon ($u=0$) can be 
        discussed in the same way. 
        It is interesing to note that for the case $Q>Gm$ no apparent horizon 
        appears in the quantum level also. 
        The black hole interior $u>0$ may appear in the classically forbidden 
        region $z^{2}<0$. 
        However, the metric variables in the region $u<0$ can evolve to $v=0$ 
        along classical trajectories without violating the WKB condition 
        $z^{2}>1$. 
        Thus no quantum penetration of the evolutionary paths into the region 
        $u>0$ occur, and no black hole state will be observed.
        
        Next, let us introduce the cosmological constant $\Lambda$ into the 
        Einstein equations. 
        Then the Lagrangian (\ref{eqn:lagrangian}) has the additional term 
    \begin{equation}
      \label{eqn:additinglambda}
          L_{\Lambda} = - \frac{V}{2 G} \Lambda \alpha v,
    \end{equation}
        and the total Hamiltonian $H_{T}$ is given by 
    \begin{equation}
         \label{eqn:desitterhamiltonian}
       H_{T} = H + \frac{\alpha V}{2 G} \Lambda v.
    \end{equation}
        Under the gauge choice $\alpha = 1$ the Schwarzschild-deSitter metric 
        becomes
    \begin{equation}
         u = - (1 - \frac{2Gm}{T} - \frac{1}{3} \Lambda T^{2}), 
                       \; \; \; \; v = T^{2}.
    \end{equation}
        The total mass $M_{T}$ defined by 
    \begin{equation}
          \label{eqn:desitterbearmass}
            M_{T} = M + \frac{\Lambda v^{3/2}}{6 G}
    \end{equation}
        is weakly commutable with $H_{T}$. 
        In this case the mass eigenstate for the Wheeler-DeWitt equation must 
        be 
    \begin{equation}
          \label{eqn:schwarzshilddesitterwavefunction}
 \Psi=N \frac{z^{p}}{(v - 2 G m v^{1/2} - (\Lambda/3) v^{2})^{p}} Z_{\pm p}(z),
    \end{equation}
        where
    \begin{equation}
          \label{eqn:schwadesiargument}
   z^{2} 
   = - \frac{V^{2} u(v - 2 G m v^{1/2} - (\Lambda/3) v^{2})}{G^{2} \hbar^{2}}.
    \end{equation}
        The wave function (\ref{eqn:schwarzshilddesitterwavefunction}) shows 
        no new quantum feature, except that if $Gm\sqrt{\Lambda}>1/3$ all the 
        region $u>0$ is classically allowed.

\newpage

\begin{center}
  Figure Captions\\
\end{center}

        Fig.1. : A time slicing indicated by the solid line in a part of the 
        Kruskal diagram of the Schwarzschild spacetime.
        The spacelike hypersurface $T$= const. is decomposed into the straight 
        line segment (A) and the curved line segment (B).\\            

	Fig.2. : The ($u,v^{1/2}$)-plane divided into four regions A, B, C and 
        D.
        (a) The classical trajectories in the regions A and D, which are drawn 
        by the solid lines corresponding to different $c$. 
        (b) The breakdown of the WKB approximation in the shaded region 
        $|z|\leq1$. 
        Any clasical trajectories must enter into this region where quantum 
        fluctuations become important, and the metric evolution can be along a 
        tunneling path between A and D.

\end{document}